\documentclass[twocolumn,notitlepage,10pt]{revtex4-1}
\usepackage{amssymb,amsmath}
\usepackage{graphicx}
\usepackage{placeins}
\usepackage{epstopdf}
\usepackage{color}
\usepackage{subfigure}
\usepackage{parskip}
\usepackage[compact]{titlesec}
\usepackage[none]{hyphenat}
\usepackage{hyperref}
\usepackage{newtxmath}

\hypersetup{
	colorlinks=true,
	linkcolor=black,
	filecolor=black,      
	urlcolor=black,
	citecolor = black,breaklinks,backref}

\usepackage{cancel}

\titlespacing{\section}{0pt}{1em}{0em}
\titlespacing{\subsection}{0pt}{1em}{0em}

\begin{document}

\title{\large Non-Hermitian tuned topological band gap}
\author{ Bikashkali Midya}
\email{midya@iiserbpr.ac.in} 
\affiliation{Indian Institute of Science Education and Research, Berhampur 760010, India}

\begin{abstract}
Externally controllable band gap properties of a material is crucial in designing optoelectronic devices with desirable properties on-demand. Here, a possibility of single parameter tuning of trivial to non-trivial topological band gap by the introduction of linear gain in an otherwise trivial insulator is investigated. Gain is selectively injected into a one dimensional lattice of dimers such that the resulting non-Hermitian Hamiltonian is symmetric under space-inversion but not under time-reversal. Inversion-symmetry of the lattice renders to probe the bulk-boundary correspondence and topological invariance by the bi-orthogonal Zak phase associated with a bulk Hamiltonian. Topological trivial to nontrivial phase transition and emergence of protected edge states are analytically shown to occur when the gain parameter is tuned across a non-Hermitian degeneracy. Tuneability of edge state location both at the boundary and inside the bulk by altering the gain distribution is discussed. Confirmation of gain-controlled topological edge state is reported in a realistic design of InGaAsP semiconductor cavity array.

\end{abstract}
\maketitle

\section{Introduction}
Topological materials, characterized by insulating bulk but unidirectionally conducting edge states appearing inside a topologically protected band gap, hold great promise for future technological applications ranging from  dissipationless electronics and photonics to quantum computers \cite{Qi2011,Kane2010,Soljacic2014,Ozawa2019}. The ability to artificially control the topology of band gaps would add a powerful degree of freedom in the design of new topological matters and devices whose properties can be tailored on demand. While the topology of a closed system is fixed because of its Hermitian character, engineered non-Hermitian interaction of energy gain and loss is an exceptional platform for  manipulating intrinsic properties of topological bands \cite{Midya2018}. 

Recent search for topological states of matter in non-Hermitian systems has revealed fundamentally novel effects \cite{Zhao2019,Schomerus2013,Lieu2018,Poli2015,Weimann2017,Esaki2011,Yao2018,Mingsen2018,Parto2018,Jin2019,Jin2017,Takata2018,Weidemann2020,Bahari2017,Kunst2018,Lee2019,Yokomizo2019,Yuce2018,Yuce2019,Longhi2019,Longhi2020,St-Jean2017,Bandres2018,Gong2018,Alvarez2018,Zhao2019,Song2019,Han2018,Zeuner2015,Leykam2017}. In particular, selective enhancement of topological interface state \cite{Schomerus2013,Poli2015}, lasing from topological edge states \cite{Bahari2017,St-Jean2017,Bandres2018}, topological light steering inside the bulk \cite{Zhao2019}, parity-time-symmetric topological interface states \cite{Weimann2017} is demonstrated. Spontaneous pumping \cite{Yuce2019}, breakdown of conventional bulk-boundary correspondence and observations of non-Hermitian skin effect \cite{Yao2018,Alvarez2018,Lee2019,Weidemann2020}  have also been reported. While topological properties of most of these non-Hermitian systems stem from their Hermitian components, it has been recently suggested \cite{Takata2018} that non-Hermiticity can also trigger topological phase transition. Topologically nontrivial phase with protected edge states has been shown to be induced by onsite gain and loss distributed inhomogeneously in a gapless photonic lattice consists of homogeneously coupled cavities.  Manipulation of both loss and gain in a heterogeneous manner, nevertheless, set new challenges in dynamic control of such non-Hermitian edge states. Contrary to acoustics \cite{Gao2020} or cold atoms in optical lattice \cite{Zoller2011,Li2019}, where loss can be tuned, in photonics loss usually accessible via material absorption \cite{Mingsen2018} or radiation due to bending \cite{Weimann2017}, which are difficult to tune.  Whereas, well developed experimental tool \cite{Zhao2019,Parto2018} of externally controllable gain in active materials provides an ideal settings for dynamic manipulation of edge states. A theoretical model, where non-Hermitian topological phase transition can be observed and corresponding edge states can be dynamically monitored by just gain tuning, is therefore, conducive for innovative device designs leveraging a successful transition for practical applications.

\begin{figure*}[ht!]
	\centering
	\includegraphics[width=0.92\textwidth]{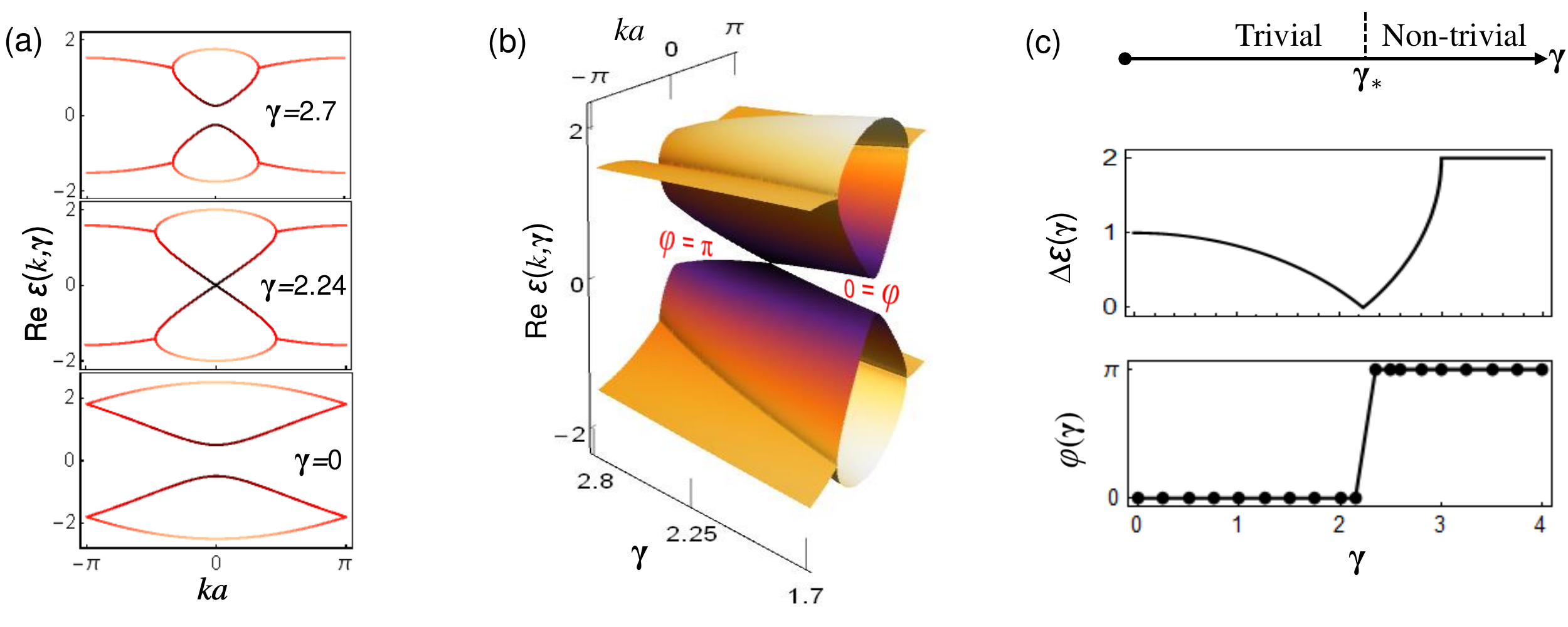}
	\caption{Non-Hermitian controlled topological band gap. (a) Real part of energy band structure is shown for selected values of gain parameter $\gamma$. (b) Real part of the band structure is shown in $k$-$\gamma$ plane when $\gamma$ is varied  in the vicinity of topological phase transition point.  (c) Energy band gap $\Delta \varepsilon$, and Zak phase $\phi$ are shown as a function of $\gamma$. Here $\beta=1.5\alpha$, $\gamma_*=2.24\alpha$, $\alpha=1$, and $\varepsilon, \gamma, \Delta\varepsilon$ are defined in the units of $\alpha$. }\label{Fig-1}
\end{figure*}

To this aim, here a possibility of topological band gap control by the introduction of non-Hermitian gain is investigated in an otherwise Su-Schrieffer-Heeger (SSH) lattice \cite{SSH,Asboth2016} in its trivial phase. Linear gain is added into the system in terms of imaginary onsite potential such that gain contrast between two nearest-neighbor sites is either $\gamma$~($\ne0$) or zero depending on whether the sites are coupled by a strong or weak bond, respectively. The resulting non-Hermitian Hamiltonian maintains space-inversion symmetry but breaks time-reversal symmetry. Inversion-symmetry of the lattice allows to investigate topological invariance and bulk-boundary correspondence by the non-Hermitian generalization of Zak phase \cite{Zak} calculated after incorporating the bi-orthogonality of eigenfunctions \cite{Lieu2018,Kunst2018,Brody2014}. Topological trivial to nontrivial phase transition and emergence of protected edge states are analytically shown to occur when the gain parameter $\gamma$ is tuned across a non-Hermitian degeneracy \cite{Miri2019}. A salient feature of the model is that gain-tuning not only allows one to generate a topological edge state, but also it provides a flexible way to alter the location of such edge state. Several possibilities of tuning the edge state location both at the boundary and inside the bulk by appropriately altering the gain distribution are shown. The proposed scheme of gain-controlled topological edge states can be experimentally verified in semiconductor laser array where linear gain can be activated either by optical illumination or carrier injection below gain saturation. Here, confirmation of gain-tuned topological edge state is reported after simulation in a realistic design of InGaAsP  microring cavity array.

\section{Gain-controlled topological band gap}
We consider a semi-analytically solvable model of four band non-Hermitian Hamiltonian in momentum ($k$-space) representation (corresponding real space realization is discussed later)
\begin{equation}
\mathcal{H}(k,\gamma)= \mathcal{H}_0(k) +  i~ \mathcal{H}_{\rm{NH}}(\gamma), \label{Eq-1}
\end{equation} 
 such that
\begin{equation}
\mathcal{H}_0(k)= \begin{bmatrix}
\beta \sigma_x & \frac{\alpha}{2}(\sigma_-+ e^{-i k a} \sigma_+)\\
\frac{\alpha}{2}(\sigma_+ + e^{i k a} \sigma_-) & \beta \sigma_x
\end{bmatrix},
\end{equation}
corresponds to the four-band generalization of Hermitian SSH model with real valued staggered coupling parameters $\beta$ and $\alpha$.  When $\beta >\alpha$, the central two bands of $\mathcal{H}_0$ are trivially gapped with no edge state [Fig.~\ref{Fig-1}(a)]. Henceforth, the ratio $\beta/\alpha$ is assumed to be greater than one, so that $\mathcal{H}_0$ is always a trivial insulator. In order to investigate the possibility of external control on closing and reopening of the central band gap, and thus inducing a topological phase transition,  we introduce non-Hermitian gain into the system characterized by the parameter $\gamma>0$ such that
\begin{equation}
\mathcal{H}_{\rm{NH}}(\gamma)=\frac{\gamma}{2}(\sigma_z\otimes\sigma_z+I).
\end{equation}
Here, $\sigma_{\pm}=\sigma_x\pm i \sigma_y$, $\sigma_{x,y,z}$ are Pauli matrices, $k\in[-\pi/a,\pi/a]$ is the Bloch momentum, and $a$ is the lattice constant. Inclusion of non-Hermiticity, in Eq.~\eqref{Eq-1}, does not violate the inversion symmetry of the system i.e. the Hamiltonian satisfies $\Pi \mathcal{H}(k,\gamma) \Pi^{-1} = \mathcal{H}(-k,\gamma)$ with respect to the inversion operator $\Pi =\sigma_x \otimes \sigma_x$, but breaks time-reversal symmetry i.e. $\mathcal{H}^*(-k,\gamma) \ne \mathcal{H}(k,\gamma)$. In the absence of  combined symmetry under parity and time-reversal, the non-Hermitian Hamiltonian $\mathcal{H}$ features complex spectrum in general \cite{Vladimir2016}. The complex band energy eigenvalues of $\mathcal{H}$ are analytically obtained as 
\begin{equation}
\begin{array}{ll}
\varepsilon_{1,4} = \pm \sqrt{\beta^2 +\alpha^2-\frac{\gamma^2}{4}+ 2\alpha\sqrt{\beta^2 \cos^2\frac{ka}{2}-\frac{\gamma^2}{4}}}+i\frac{\gamma}{2}\\\\
\varepsilon_{2,3} = \pm \sqrt{\beta^2 +\alpha^2-\frac{\gamma^2}{4}- 2\alpha\sqrt{\beta^2 \cos^2\frac{ka}{2}-\frac{\gamma^2}{4}}}+i\frac{\gamma}{2},
\end{array}
\end{equation}
which are shown in Fig.~\ref{Fig-1}(a) for specific values of $\gamma$, and in Fig.~\ref{Fig-1}(b) when the parameter $\gamma$ was continuously varied near the physically interesting regime. Remarkably, initially opened central band gap is seen to be closed and reopened at $k=0$ when  $\gamma$ attains a threshold value at which both the real and imaginary (not shown here) parts of the central two bands coalesce. The threshold value of $\gamma$ leading to such non-Hermitian degeneracy can be predicted from the analytical expression of the energy gap between second and third bands, which is given by 
\begin{equation}
\Delta\varepsilon (\gamma) = \left|\mbox{Re}\left(\sqrt{4 \beta^2-\gamma^2}\right)-2\alpha\right|
\label{Eq-gap}
\end{equation}
at $k=0$. This shows that for fixed values of $\beta$ and $\alpha$, the band gap is a function of the gain parameter $\gamma$, and hence can be tuned by changing $\gamma$. Indeed, as shown in Fig.~\ref{Fig-1}(c), $\Delta\varepsilon$ monotonically decreases as $\gamma$ is increased from zero, and eventually vanishes when $\gamma$ attains critical value $\gamma_*~=~2\sqrt{\beta^2-\alpha^2}$. The vanishing of band gap at $(k,\gamma)=(0,\gamma_*)$ is also associated with the exceptional point degeneracy  at which  two band energies $\varepsilon_{2}$ and $\varepsilon_{3}$ coalesce to a single value equal to $i\gamma_*/2$.  For $\gamma_*<\gamma<2\beta$, the band gap increases monotonically and attains a maximum value equal to $2\alpha$ which remains unchanged for $\gamma\ge2\beta$. Note that a cusp like singularity is observed in the vicinity of gap closing point, $\gamma=\gamma_*$, which separates two distinct parameter regimes e.g. $\gamma<\gamma_*$ and $\gamma>\gamma_*$ [Fig.~\ref{Fig-1}(c)].

The band gap given in Eq.~(\ref{Eq-gap}) can also be represented as $\Delta\varepsilon=2|\beta_{\rm{eff}}-\alpha|$ in terms of an effective parameter
\begin{equation} \beta_{\rm{eff}}(\gamma)=\rm{Re}\left(\sqrt{\beta^2-\gamma^2/4}\right).\label{Eq-effective-beta}
\end{equation} 
This effective representation resembles the well know band gap relation in a conventional two-band Hermitian SSH model \cite{Asboth2016}, provided the strong coupling parameter $\beta$ is replaced by the effective one i.e. $\beta_{\rm eff}$. The relation (\ref{Eq-effective-beta}) shows that $\beta_{\rm{eff}}$ reduces from its maximum value $\beta$ to a minimum value $0$, when the gain parameter increases from $\gamma=0$ to $\gamma=2\beta$. Whereas the band gap $\Delta\varepsilon$ vanishes when $\beta_{\rm{eff}}=\alpha$ is satisfied at $\gamma=\gamma_*$, and reopens again when $\beta_{\rm{eff}}<\alpha$ i.e. when $\gamma>\gamma_*$. A topological trivial or a nontrivial phase may therefore occurs depending on whether the effective coupling is greater or less than the weak coupling (i.e.  $\beta_{\rm{eff}}>\alpha$ or $\beta_{\rm{eff}}<\alpha$), respectively. This, in turn, indicates to the intriguing possibility of triggering topological phase transition enabled by the effective control of one of the coupling parameters. 

\begin{figure}[ht!]
	\centering
	\includegraphics[width=0.49\textwidth]{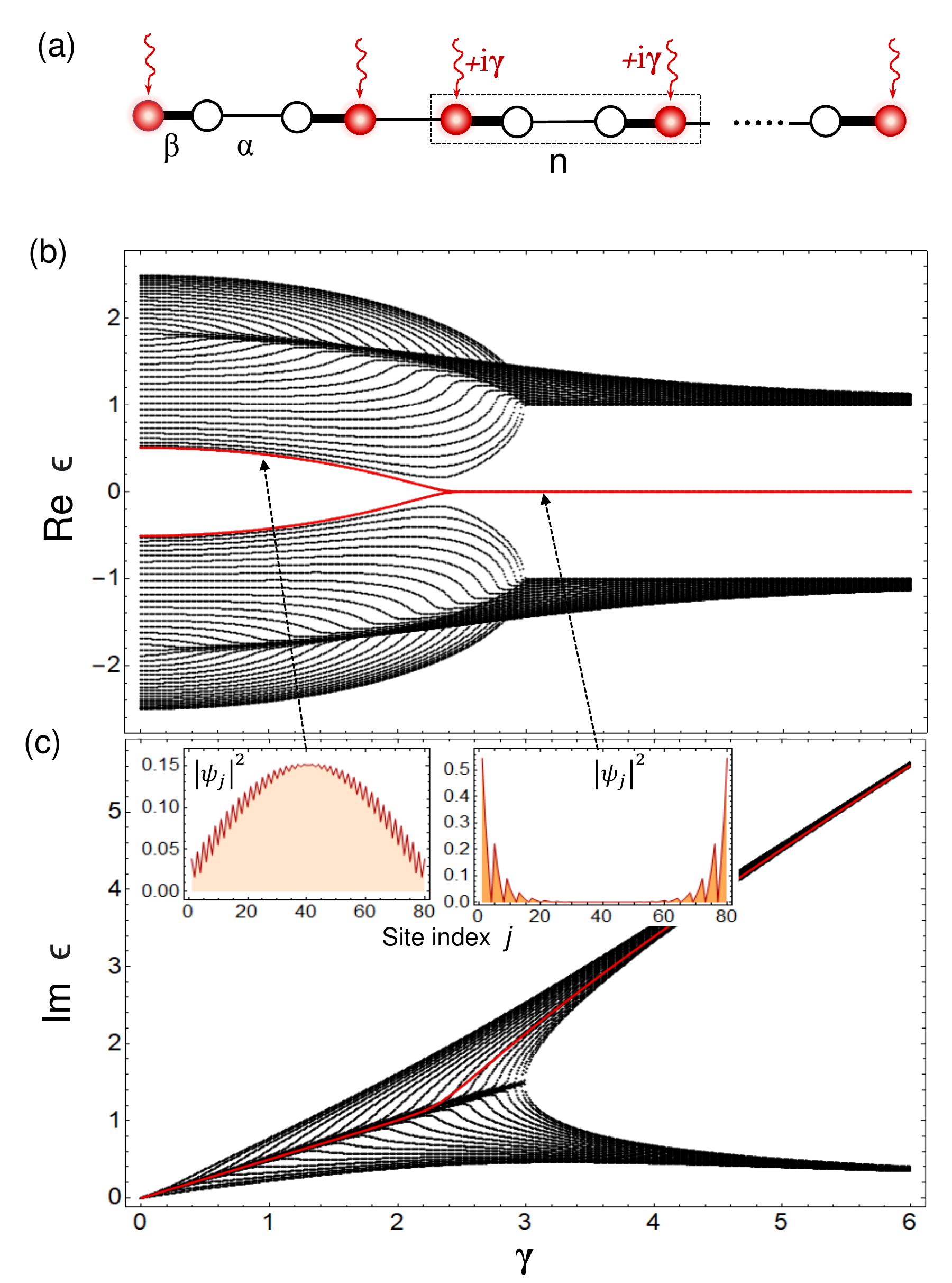} 
	\caption{Gain-induced topological edge states. (a) Shows the schematic of a non-Hermitian lattice with staggered coupling $\beta$ and $\alpha$. Distributed gain is introduced into the sites shown in red color, whereas empty circles denote sites with zero onsite potential. The unit cell is shown by the dashed rectangle. (b) Real and (c) imaginary parts of the spectrum are shown when gain parameter $\gamma$ is varied (in the units of $\alpha$) across the non-Hermitian degeneracy. The evolution of mid gap states are highlighted in red color. A pair of zero energy edge states emerges for $\gamma >\gamma_* =2.24\alpha$. Unnormalized intensity distribution $|\psi_j|^2$ corresponding to one of the mid gap states, before and after topological phase transition, are shown in the inset of lower panel. Here $4N=80$, $\beta=1.5\alpha$ and  $\alpha=1$. }\label{Fig-2}
\end{figure}

These results are in stark contrast to the parity-time (PT) symmetric generalized two-band SSH model \cite{Lieu2018}, where  intrinsic topology of the corresponding Hermitian band remains unaltered by the introduction of onsite gain and loss. The PT-symmetric system usually becomes gapless, when gain/loss parameter exceeds the PT-breaking threshold; closing of bandgap, however, does not destroy the existing edge state. Contrary, in the four-band generalized non-hermitian SSH model, studied here, the energy gap of the system closes and reopens by non-Hermitian symmetry breaking at an exceptional point, which not only allows one to alters the topology of the system but also enables to create new edge states not existing in the Hermitian limit. This hallmark feature is used below to control the non-Hermiticity induced edge states.

 Above mentioned non-Hermiticity enabled band-gap closing and re-opening phenomena is one of the signature of topological phase transition. In order to probe such phase transition,  here we quantify the topological invariance of the center band gap by the Zak phases (mod $2\pi$) summed over for all occupied bands below the gap \cite{Xiao2014,Midya2018a}:
\begin{equation}
\varphi(\gamma)=\sum\limits_{n=1}^2 i \int_{-\pi/a}^{\pi/a}  u_n(k,-\gamma)^* \frac{d}{dk} u_n(k,\gamma)~dk,
\end{equation} 
where bi-orthogonal scalar product \cite{Kunst2018,Lieu2018} has been used. Here $u_n(k,\gamma)$ are the right-eigenvectors of $\mathcal{H}$ associated to the band energies $\varepsilon_n(k)$, and $u_n(k,-\gamma)$ are the corresponding left-eigenvectors of $\mathcal{H}$ which follows from the fact that $\mathcal{H}(k,\gamma)^\dag=\mathcal{H}(k,-\gamma)$. Indeed, numerically computed Zak phase, shown in Fig.~\ref{Fig-1}(c), confirms the existence of two topologically distinct parameter regimes: trivial ($\varphi=0$) when $\gamma~<~\gamma_*$, and nontrivial ($\varphi=\pi$) for $\gamma~>~\gamma_*$. According to bulk-boundary correspondence, therefore, topologically protected localized states can emerge either at the open boundaries of a non-trivial lattice or at the interface between a trivial and a non-trivial lattices. Such possibilities are discussed below.

\section{Gain-induced tunable edge states}
A specific realization of the above mentioned Bloch Hamiltonian $\mathcal{H}$ and the corresponding topology can be achieved by the real space Hamiltonian $H=H_0 + i~ H_{\rm{NH}}$  [shown in Fig.~\ref{Fig-2}(a)], such that 
\begin{equation}\begin{array}{ll}
H_0=\sum\limits_{n=1}^{N}&\left(\beta |n,1\rangle \langle n,2|+  \alpha |n,2\rangle \langle n,3| + \beta |n,3\rangle \langle n,4|\right.\\
&\left. + \alpha |n,4\rangle \langle n+1,1| + h.c. \right),
 \end{array}
\end{equation}
represents a conventional SSH model with four elements in a unit cell, and
\begin{equation}
H_{\rm{NH}}= \gamma \sum\limits_{n=1}^{N} \left(|n,1\rangle \langle n,1|+ |n,4\rangle \langle n,4| \right)
\end{equation}
represents onsite gain distribution. Here first and second indices in the position basis $\{|n,m\rangle\}$ enumerate unit cells and sites inside a unit cell, respectively. The lattice contains total $4N$ sites, $N$ unit cells and four sites in each of the unit cells. Gain is selectively introduced into the first and fourth lattice sites inside a unit cell such that gain contrast between two nearest-neighbor sites attached to a strong bond is $\gamma$, while that between weak bond is zero. The stationary solution of the corresponding Schr\"odinger equation $i\frac{\partial |\psi\rangle}{\partial t} = H |\psi\rangle$ can be sought in the form $|\psi\rangle = e^{-i \epsilon t} \sum\limits_{n,m} \psi_{n,m} |n,m\rangle$, such that  the amplitudes $\psi_{n,m}$ and energy $\epsilon$ satisfy
\begin{equation}
\begin{array}{lll}
\alpha \psi_{n-1,4} + \beta \psi_{n,2} +i \gamma \psi_{n,1} = \epsilon \psi_{n,1} \\
\beta \psi_{n,1} +\alpha \psi_{n,3} = \epsilon \psi_{n,2} \\
\alpha \psi_{n,2} +\beta \psi_{n,4} = \epsilon \psi_{n,3} \\
\beta \psi_{n,3} + \alpha \psi_{n+1,1} + i \gamma \psi_{n,4} = \epsilon \psi_{n,4},
\end{array}
\end{equation}
where $n=1,2,\cdots, N$. Above set of equations has been solved for a finite number of lattice sites ($4N=80$) after imposing open boundary conditions. Real and imaginary parts of the corresponding discrete spectrum as well as the intensity distribution of the midgap state are shown in Fig.~\ref{Fig-2}(b) and (c) for specific values of $\beta$ and $\alpha$ when $\gamma$ was varied (the site index $(n,m)$ has been reset to $j\equiv4(n-1)+m$ for convenient presentation). As predicted in previous section, two zero-energy degenerate edge states are seen to emerge for $\gamma >\gamma_*$ inside the topologically non-trivial band gap.  On the other hand, the band gap is completely empty when $\gamma<\gamma_*$ i.e. the trivial phase. These results are in complete agreement with the bulk-boundary correspondence, which holds here although the system is not Hermitian.

\begin{figure}[t!]
	\centering
	\includegraphics[width=0.49\textwidth]{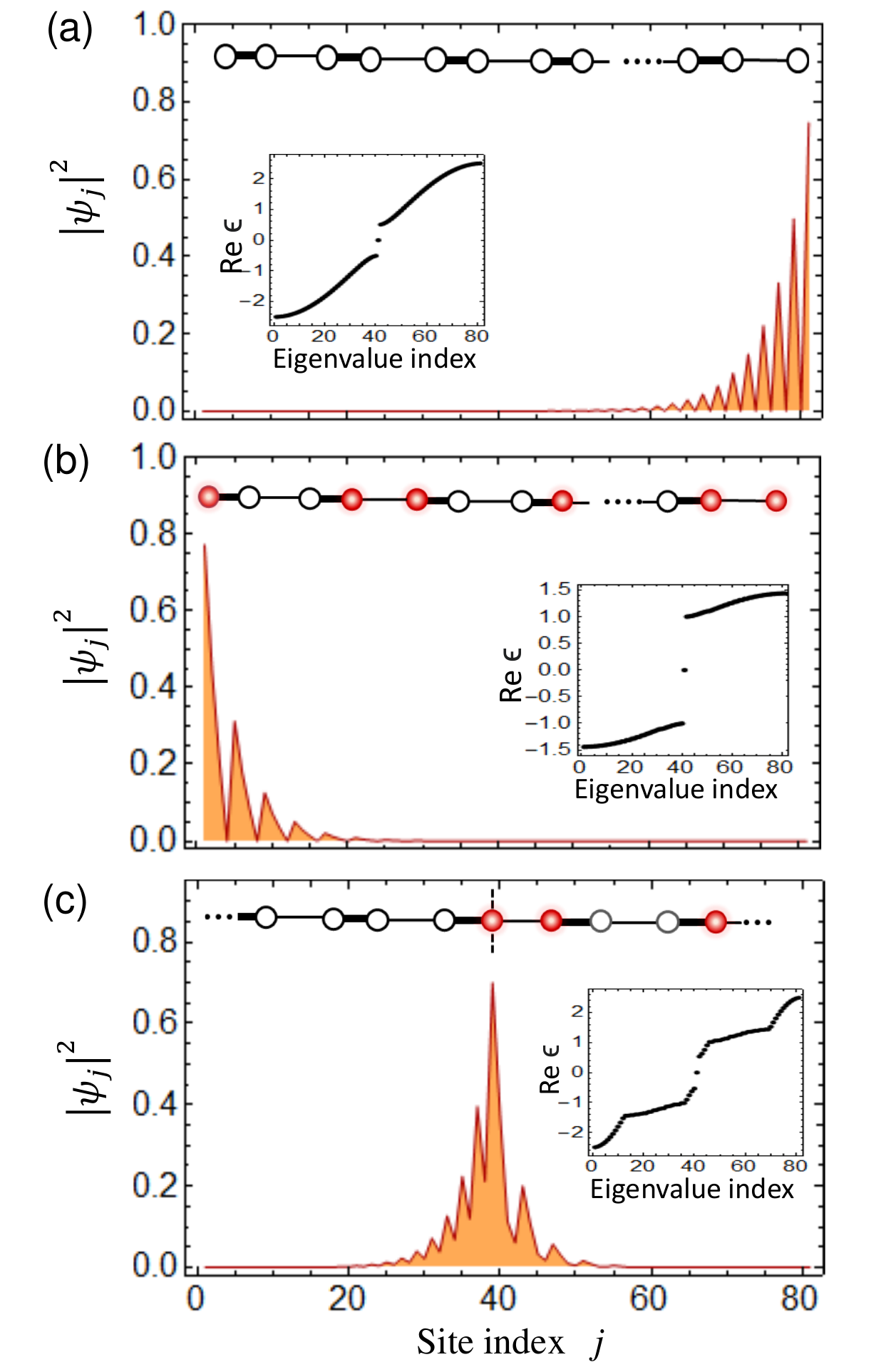} 
	\caption{Spatially tunable edge states. (a) Edge state and corresponding spectrum of a lattice with $(4N+1)=81$ sites without gain in the system. (b) Switching of the edge state location to the left is shown when gain is introduced into the entire lattice. (c) When gain is applied into a part of the lattice, the topological state appears at the interface between two semi lattices, one with and other without gain. The real part of the corresponding spectrum with zero energy mode is shown in the inset of each panels. Here  $\beta=1.5\alpha$, $\gamma_*=2.24\alpha$, $\gamma=3\alpha$, and $\alpha=1$. }\label{Fig-3}
\end{figure}

Few remarks on the gain-induced edge states are in order here. First remark is concerned with the robustness of the edge states against certain disorder.  For fixed values of coupling constants $\beta$ and $\alpha$, the edge states created here are robust against random onsite gain disorders of strength $\Delta\gamma_{n,m}~=~\gamma-\gamma_{n,m}$ satisfying $\Delta\gamma_{n,m}\le\Delta\varepsilon$, for $m=1,4$ and $n=1,\cdots, N$. An example of such robustness against onsite non-Hermitian disorder is given in Fig.~\ref{Fig-3}(b). Second remark is related to dynamical stability of the edge states at larger time. Unlike Hermitian edge states which are associated with real spectrum, the spectrum of non-Hermitian edge states is complex [Fig.~\ref{Fig-2}(b) and (c)]. Here, the positive imaginary part of the energy is associated with the rate of amplification of the corresponding state. In a multi-state system, the state with largest gain grows rapidly relative to other states and becomes dominant at large times. In the present context, the edge state is not dominating at larger time because some of the bulk states have gain exceeding that of an edge state. This is clear from the imaginary part of the spectrum shown in Fig.~{\ref{Fig-2}(b)}, where corresponding gain for the edge states are highlighted in red color. Nevertheless, the gain for an edge state can be made maximum by increasing onsite gain only at the edge of the lattice (this conclusion is verified in numerical simulation). As the bulk remains unaffected, such a gain bias, only at the edge, does not alter the underlying bulk topology. Further details regarding dynamical stability is not studied here.

	\begin{figure*}[t!]
	\centering
	\includegraphics[width=0.85\textwidth]{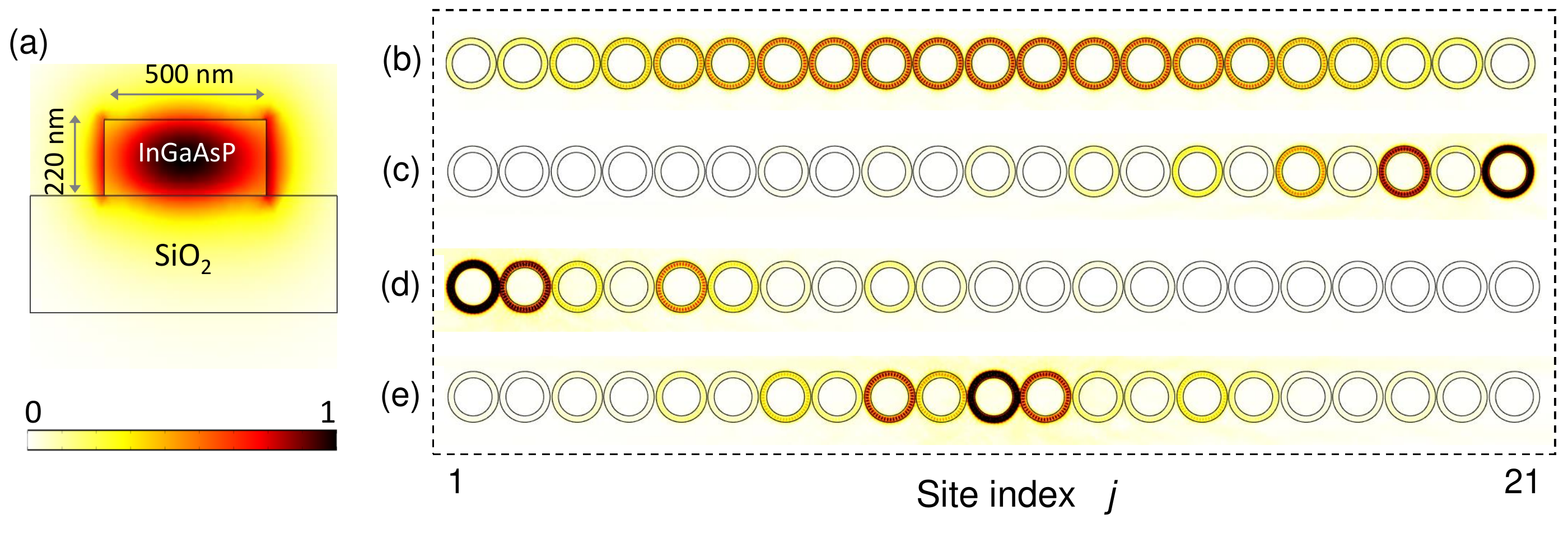} 
	\caption{Observation of gain-controlled topological edge state in a microring cavity array configured according to Fig.~\ref{Fig-3}. (a)~Cross section of a cavity showing the mode confinement inside the gain medium. A bulk supermode in (b), and three topological edge supermodes in (c), (d) and (e), are obtained after simulation in an array of $21$ micriring cavities placed in alternating distances of 150~nm and 240~nm apart. (b) and (c) were obtained in the absence of gain in the system. In (d) and (e), $\gamma=240$GHz was considered in selected rings according to Fig.~\ref{Fig-3}(b) and (c), respectively. The wavelength of bulk mode corresponds to 1505~nm, whereas that of edge modes is $1508$~nm.}\label{Fig-4}
\end{figure*}

A fascinating feature of our model is that the edge state properties can flexibly be controlled solely by the gain parameter. By altering the gain distribution appropriately it is possible to induce an edge state at an arbitrary location in the lattice. To illustrate this possibility we consider a lattice of $4N+1$ sites, so that the lattice starts with a strong bond at the left and terminates with a weak bond at the right. In this case a single topological zero-energy state appears at the right edge even in the absence of gain [as shown in the Fig.~\ref{Fig-3}(a)]. The edge state location can be switched to the left edge by the application of gain whose strength is greater than its threshold value. This situation is shown in Fig.~\ref{Fig-3}(b). In this case, the nature of strong-weak-strong-weak coupling configuration of the Hermitian lattice is altered to effectively weak-strong-weak-strong configuration. As a consequence, a new zero-energy edge state appears at the left edge which starts with an effectively weaker bond. Note that, a random gain disorder $|\Delta\gamma_{n,m}|<0.2\gamma$ is introduced in numerical simulations to test the robustness of the edge state in Fig.~\ref{Fig-3}(b). The edge state can also be created inside the bulk at an interface between two semi-lattices, one with gain and another without gain, as shown in Fig.~\ref{Fig-3}(c).  Here the semi-lattice without gain is topologically trivial whereas the one with gain is nontrivial. Edge state, therefore, appears at the interface of two different topological lattices. Thus, contrary to Hermitian topological system, where edge state location is fixed, here edge state can be dynamically controlled by gain tailoring.

\section{Verification of tunable edge states in a microcavity array}
Above mentioned theoretical prediction of tunable edge states have been confirmed by simulation in an array of $21$ InGaAsP semiconductor microring cavities placed in alternating distances on a SiO$_2$ (refractive index = 1.44) substrate. The simulation were performed in COMSOL multi-physics using the full-wave finite element method. Geometry of each ring have inner and outer radii of 2.5~$\mu$m and 3~$\mu$m, respectively, and a height of 220~nm. Imaginary part of the complex refractive index $n=n'+i n''$ of InGaAsP was tuned for controlling gain in the system. While the real part $n' =3.4$ is fixed around the wavelength $1508$~nm (which is equal to the  resonant wavelength of an isolated cavity), the parameter $n''$ can be activated and changed flexibly in experiments by pumping the cavities either optically or electrically \cite{Miidya2019}. 
In order to verify the results presented in Fig.~\ref{Fig-3}, the alternating 150~nm and 240~nm edge-to-edge separations between adjacent rings were set for the strong coupling $\beta=120$~GHz and the weak coupling $\alpha=80$~GHz, respectively. Furthermore, the parameter $n''=0.003$ (or $n''=0$) was considered to achieve required gain  $\gamma=3\alpha$ (or $\gamma=0$) in respective cavities. The simulation result is shown in Fig.~\ref{Fig-4}. In the absence of gain in the entire array, a conventional localized state is observed at the right edge [Fig.~\ref{Fig-4}(c)]. When gain is introduced into the system, the edge state shifted either to the left edge [Fig.~\ref{Fig-4}(d)] or to the center  [Fig.~\ref{Fig-4}(e)] of the array, depending on the gain distributions. A gain offset $\Delta \gamma=40$~GHz (corresponding $\Delta n''=5\times10^{-4}$) was added into a randomly selected site in the bulk to verify the robustness of the edge state presented in Fig.~\ref{Fig-4}(d). Finally, the `zero energy' of edge states is verified by the fact that all the three edge supermodes have a wavelength equal to 1508~nm which is identical to the individual cavity resonance in isolation.

To conclude, in this paper a possibility of topological band gap tuning by the introduction of optical gain into a trivially gapped insulator is discussed. Trivial to non-trivial topological phase transition and emergence of topologically protected edge states are shown to occur when the gain parameter exceeds a threshold value at which central two bulk bands coalesce due to a non-Hermitian degeneracy. A notable  feature of the model is that by simply changing the gain distribution it is possible to arbitrarily alter the spatial location of the edge state. Theoretical predictions is also confirmed here in a realistic design of InGaAsP microcavity array. The theory of gain-tailored edge state is thus promising for future experiments on active semiconductor materials where gain can be injected by optical illumination or electrical pumping, and offers novel possibilities for  functional topological devices, particularly in tunable laser applications where signal can be amplified in desired channels. As a final remark, the theory is equally valid if one replaces gain by dissipation which can be tuned in other physical settings e.g. in acoustics or cold atoms in optical lattice.

\end{document}